\documentclass[english,prl, twocolumn, showpacs]{revtex4}
\usepackage[T1]{fontenc}
\usepackage[latin9]{inputenc}
\usepackage{color}
\usepackage{graphicx}
\usepackage{amssymb}

\makeatletter
\usepackage{color}
\usepackage{bm}
\def\kbar{{\mathchar'26\mkern-9mu k}}

\makeatother

\usepackage{babel}

\begin{document}

\title{Tracking quasi-classical chaos in ultracold boson gases}

\author{Maxence Lepers}

\author{V\'eronique Zehnl\'e}

\author{Jean Claude Garreau}

\affiliation{Laboratoire de Physique des Lasers, Atomes et Mol\'ecules, Universit\'e
des Sciences et Technologies de Lille; CNRS; F-59655 Villeneuve d'Ascq
Cedex}

\homepage{http://phlam.univ-lille1.fr/atfr/cq}

\begin{abstract}
We study the dynamics of a ultra-cold boson gas in a lattice submitted
to a constant force. We track the route of the system towards chaos
created by the many-body-induced nonlinearity and show that relevant
information can be extracted from an experimentally accessible quantity,
the gas mean position. The threshold nonlinearity for the appearance
of chaotic behavior is deduced from KAM arguments and agrees with
the value obtained by calculating the associated Lyapunov exponent. 
\end{abstract}

\pacs{03.75.Nt, 05.45.Ac, 37.10.Jk}

\maketitle
Recent advances in the physics of cold atoms paved the way for the
investigation of fundamental quantum problems with unprecedented cleanness.
{}``Quantum chaos'' is one of the most fascinating among these problems,
because in such case the correspondence between the dynamics of the
quantum system and its classical counterpart is nontrivial. The reasons
for that are twofold: First, quantum particles obey Heisenberg inequalities,
hence their dynamics cannot be described in terms of phase-space trajectories.
Second, sensitivity to initial conditions is not observed, as the
Schrödinger equation is linear. For these reasons, \emph{quantum chaos}
has often been defined as {}``the behavior of a quantum system whose
classical limit is chaotic''. Whereas this is a reasonable definition,
it is clear that the actual quantum dynamics has no direct relation
to classical chaos. Much work thus concentrates in finding {}``reminiscences''
of classical chaos that might survive in the quantum system, the so-called
{}``signatures'' of quantum chaos \citep{HaakeQSC}.

The realization in 1995 of the first Bose-Einstein condensate with
laser-cooled atoms \citep{CornellWieman:BECFirst:S95,Hulet:BECFirst:PRL95,Ketterle:Nobel:RMP02}
opened a new way for investigating (truly) nonlinear dynamics in quantum
systems. In an {}``ideal'' (i.e.~in the zero temperature limit)
Bose-Einstein condensate (BEC), the atoms are indistinguishable and
form a mesoscopic object which can be described by a {}``collective''
single wavefunction. The BEC's dynamical behavior is then described
by the Gross-Pitaevskii equation which includes a \emph{nonlinear}
term due to atomic interactions \citep{Stringari:BECRevTh:RMP99}.
The solutions of such equation can -- and do -- present sensitivity
to initial conditions, leading to {}``classical-like'' instabilities,
a possibility that attracted much attention from both theoreticians
\citep{AP:ChaosBEC:PRL03,Javanainen:InstabilityBEC:PRL04,Smerzi:InstabilityBEC:PRL04,
Hai:WSChaosBEC:PB05,Korsch:BlochOscBEC:PRE05,Niu:InstabilityBEC:PRA06,
Wimberger:TunnelingNonlinearWS:JPB06,VanNoort:BECPeriodicLat:NS07,Tania:KickedBEC:PRA08}
and experimentalists \citep{Inguscio:InstabilityBEC:PRL04,Arimondo:InstabilityBEC:OE08}. 

\begin{figure}
\begin{centering}
\includegraphics[width=7cm]{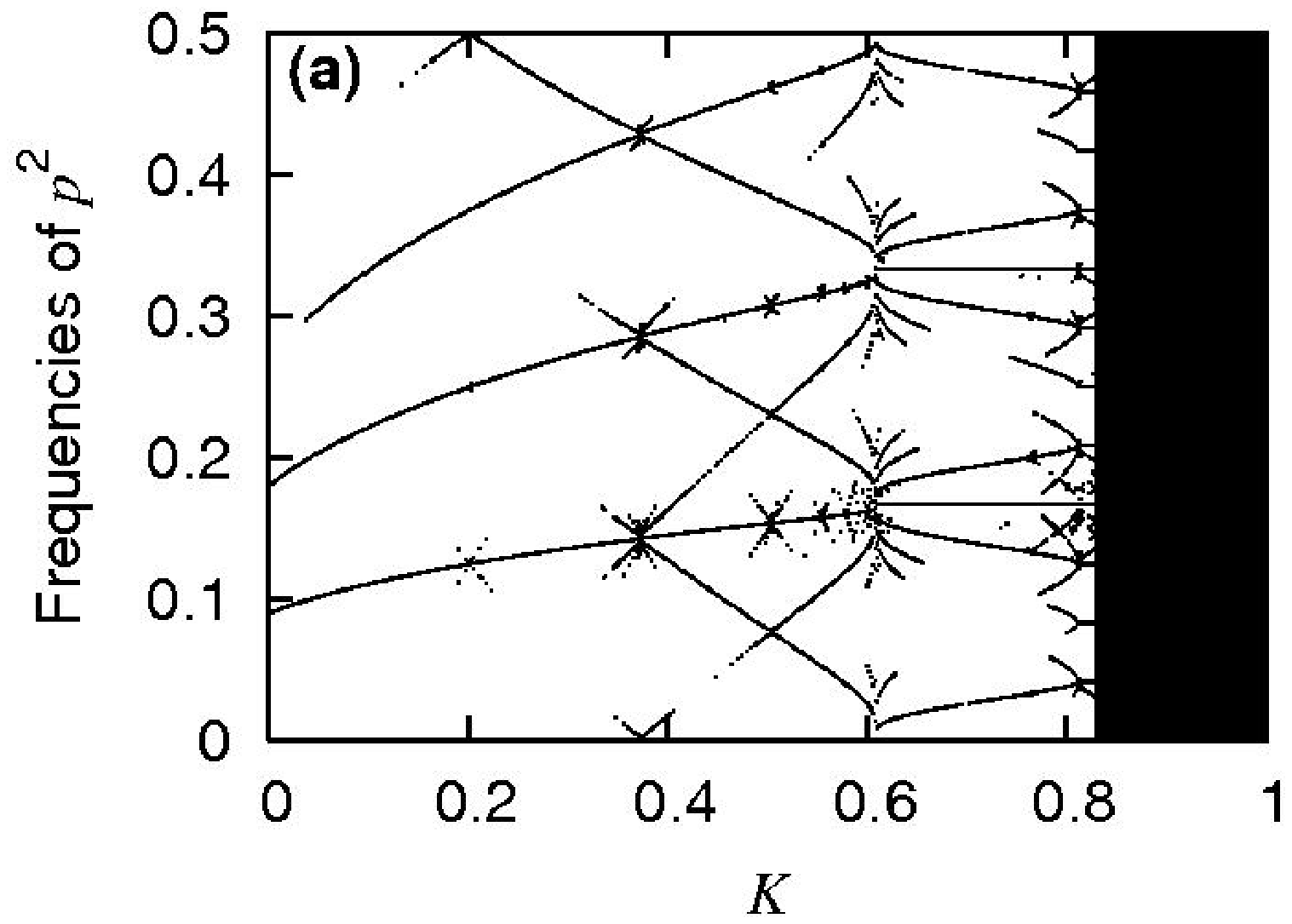}
\par\end{centering}

\begin{centering}
\includegraphics[width=8cm]{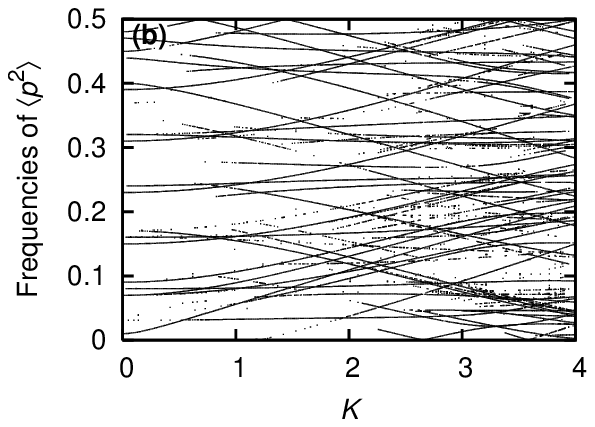}
\par\end{centering}

\caption{\label{fig:QKR-CKR}Classical and quantum dynamics of the kicked rotor.
For each value of $K$, we numerically calculate $p^{2}(t)$ (classical
case) and $\left\langle p^{2}(t)\right\rangle $ (quantum case) up
to 2000 kicks, perform the Fourier transform and pinpoint the frequencies
whose Fourier amplitude is above a threshold (here 1/500 of the maximum
amplitude). (a) Frequencies in the classical dynamics: A dense spectrum
appears for $K\gtrsim0.82$ (initial conditions $x=0$ and $p=0.56$).
(b) Frequencies in the quantum dynamics: The spectrum is complex,
but always discrete ($\kbar=2.89$, the initial state is a Gaussian
in momentum space of FWHM $5\kbar$ centered at $p=0.56$).}

\end{figure}

In order to make these ideas clear, let us first consider a simpler
system, the kicked rotor, whose experimental realization with laser
cooled atoms \citep{Raizen:QKRFirst:PRL95} lead to an impressive
burst of experimental work \citep{Amman:LDynNoise:PRL98,DArcy:AccModes:PRL99,
Philips:QuantumResTalbot:PRL99,AP:Bicolor:PRL00,AP:SubFourier:PRL02,
Leonhardt:KREarlyTimeDiff:PRL04,AP:Reversibility:PRL05,
Monteiro:LocDelocDoubleKick:PRL06}.
This system is formed by a particle periodically ``kicked'' by
a sinusoidal force. The classical dynamics is characterized by a single
parameter $K$, which is the normalized amplitude of the potential
\citep{Raizen:QKRFirst:PRL95}, whereas the {}``quanticity'' of the
quantum system is described by a {}``normalized Planck constant''
$\kbar$ (the dynamics become classical as $\kbar\rightarrow0$). 

In order to characterize the dynamics, one may chose, for instance,
the particle's square momentum as the relevant quantity ($p^{2}(t)$
in the classical case, $\left\langle p^{2}(t)\right\rangle $ in the
quantum case). By taking the Fourier transform of the above quantity
and pinpointing the frequencies appearing in the evolution of the
system for different values of $K$ and fixed initial conditions,
we can characterize in a common language the classical and the quantum
versions. The result is plotted in Fig.~\ref{fig:QKR-CKR}. 

The classical dynamics, Fig.~\ref{fig:QKR-CKR}a, shows resonances
(or frequency-lock events) each time a frequency become very close
to another one. For $K\gtrsim0.8$, the frequencies in the spectrum
{}``coalesce'' to form of a dense spectrum, which is a signature
of a chaotic dynamics. In the quantum case, Fig.~\ref{fig:QKR-CKR}b,
the spectrum is richer and present numerous frequency crossings, but
no dense spectrum is observed.

Let us now consider a one-dimensional BEC of particles of mass $M$
placed in a sinusoidal lattice formed by counterpropagating laser
beams of wavelength $\lambda_{L}$, and subjected to a linear force.
The system is described by the Gross-Pitaevskii equation (GPE) 
\begin{equation}
i\frac{\partial\psi(x,t)}{\partial t}=\left(\frac{P^{2}}{2m}+V_{0}\cos(2\pi x)
+Fx+g\left|\psi\right|^{2}\right)\psi(x,t),
\label{eq:GPE}
\end{equation}
with lengths measured in units of the lattice step $d=\lambda_{L}/2$,
energy measured in units of the recoil energy $E_{R}=\hbar^{2}k_{L}^{2}/(2M)$
(with $k_{L}=2\pi/\lambda_{L}$), time in units of $\hbar/E_{R}$,
the force $F$ in units of $E_{r}/d$, $P=-i\partial/\partial x$,
the reduced mass is $m=\pi^{2}/2$ and $g$ is the (1D) nonlinear parameter.
The eigenstates of the linear part of the Hamiltonian (\ref{eq:GPE}),
obtained by setting $g=0$, are the so-called Wannier-Stark (WS) states
\citep{Wannier:WS:PR60,AP:WannierStark:PRA02,AP:ChaosBEC:PRL03}.
In order to simplify the discussion, let us we suppose that the BEC
energy is low enough that its wavefunction $\psi$ can be expanded
only on the lowest-energy WS state for each potential well, hereafter
noted as $\varphi_{n}(x)$ (corresponding to the $n^{\mathrm{th}}$
well). These states are centered at each potential well, obey the
symmetry relation $\varphi_{n+m}(x)$$=\varphi_{n}(x-m)$, and form
a ladder of eigenenergies $E_{n}=n\omega_{B}$, where $\omega_{B}=Fd/\hbar$
($F$ in our normalized units) is the Bloch frequency. Thus, putting
$\varphi(x)\equiv\varphi_{0}(x)$
\begin{equation}
\psi(x,t)=\sum_{n}\sqrt{I_{n}}e^{-i\theta_{n}}\varphi(x-n),
\label{eq:psi0}
\end{equation}
where the eigenstates population $I_{n}$ and phase $\theta_{n}$
are real functions of the time.

Following the approach of \citep{AP:ChaosBEC:PRL03}, we insert Eq.~(\ref{eq:psi0})
in Eq.~(\ref{eq:GPE}), and obtain Hamilton equations for the populations
and phases. The Hamiltonian then appears as a sum of an integrable
part $H_{0}$ and a non-integrable perturbation $H_{1}$:
\begin{equation}
H=H_{0}(I)+\epsilon H_{1}(I,\theta),
\label{eq:KAMperturbation}
\end{equation}
with $\epsilon\propto g\chi_{01}$, where $\chi_{0i}\equiv\int dx\varphi^{3}(x)\varphi(x-i)$
is a measure of the superposition of eigenfunctions corresponding
to different wells. For small $\epsilon$, the system is quasi-integrable
and fits into the general frame of the KAM theorem \citep{Lichtenberg:ChaoticDynamics:82}.
Therefore, the quantum-coherent evolution of a BEC displays classical-like
KAM-structured chaos. We call this, ``quasi-classical'' chaos.

If we temporarily neglect the nonlinear term in Eq.~(\ref{eq:GPE}),
we see that the phases evolve as $\theta_{n}(t)=\theta_{n0}+n\omega_{B}t$.
To the first order in $\epsilon$, the effect of the nonlinearity
is to introduce a population-dependent correction to this phase, producing
a harmonic evolution with frequencies $\omega_{n}=n\omega_{B}+g\chi_{00}I_{n}$.
The intensities are constants of motion, $I_{n}(t)=I_{n}(t=0)$ and
the ``self-coupling'' parameter $\chi_{00}$ depends only $V_{0}$
and $F$. The BEC dynamics is thus governed by the Bohr frequencies
\begin{equation}
\omega_{nm}\equiv\omega_{n}-\omega_{m}=(n-m)\omega_{B}+g\chi_{00}(I_{n}-I_{m}).
\label{eq:phasediff}
\end{equation}
In order to simplify further our description, let us restrict the
dynamics to three adjacent potential wells, i.e.~we set $I_{n}\equiv0$
if $n\neq-1,0,1$ in Eq.~(\ref{eq:psi0}). The system is then four-dimensional,
the dynamical variables being two populations (since $I_{-1}+I_{0}+I_{1}=1$)
and two relative phases. 

\begin{figure}
\begin{centering}
\includegraphics[width=8cm]{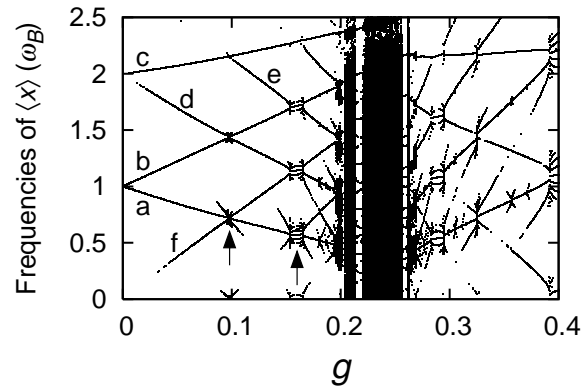}
\par\end{centering}

\caption{\label{fig:Frequencies} Frequencies present in the spectrum 
of $\left\langle x(t)\right\rangle $
\emph{vs} $g$ (the threshold is 1/100 of the maximum amplitude).
For small $g$, the Bohr frequencies {[}cf. Eq.~(\ref{eq:phasediff})]
are $\omega_{10}=\omega_{B}-0.4g\chi_{00}$, $\omega_{0-1}=\omega_{B}+0.55g\chi_{00}$
and $\omega_{1-1}=2\omega_{B}+0.15g\chi_{00}$, corresponding to branches
a, b, and c, respectively. Harmonics of $\omega_{10}$ are also seen
(branches d and e) and $\Omega_{1,-1,0}$ (branch f). For $0.22\lesssim g\lesssim0.26$
one observes dense-spectrum windows corresponding to quasi-classical
chaos. Parameters are $\omega_{B}=0.25$, $V=5$ ($\chi_{00}=2$);
initial conditions are $I_{0}=0.65$, $I_{1}=0.25$, $\theta_{0}=\theta_{-1}=0$,
$\theta_{1}=\pi$. }

\end{figure}

The onset of chaos in this system is shown in Fig.~\ref{fig:Frequencies},
which is the equivalent of Fig.~\ref{fig:QKR-CKR} for the BEC dynamics,
with the difference that we used the average position $\left\langle x(t)\right\rangle $
instead of $\left\langle p^{2}(t)\right\rangle $. Strikingly, the
plot resembles more closely to the \emph{classical} (Fig.~\ref{fig:QKR-CKR}a)
than to the quantum (Fig.~\ref{fig:QKR-CKR}b) dynamics of the kicked rotor. 

We can interpret the main features of Fig.~\ref{fig:Frequencies}
from simple arguments. At $g=0$ there are only three Bohr frequencies
in the model: Two of them are degenerated, $\omega_{0-1}$ $=\omega_{10}$
$=\omega_{B}$, and correspond to energy difference between neighbor
wells, whereas $\omega_{-11}=2\omega_{B}$ corresponds to next-to-neighbor
energy difference. The low-$g$ structure can be understood using
Eq.~(\ref{eq:phasediff}). The frequencies present around $g=0.05$,
for instance, in Fig.~\ref{fig:Frequencies} are of the form 
\begin{equation}
\Omega_{pqr}=p\omega_{0-1}+q\omega_{10}+r\omega_{1-1}
\label{eq:harmonics}
\end{equation}
with $p,q,r$ integer. Namely, from top to bottom, we found $\omega_{1-1},$
$2\omega_{10},$ $\omega_{0-1},$ $\omega_{10}$ and $\omega_{0-1}-\omega_{10}$.
These frequencies are only weakly perturbed by the nonlinearity, which
manifests itself by the slight curvature of the branches. A KAM-type
expansion in powers of $\epsilon$ shows that the weight of a frequency
$\Omega_{pqr}$ in the spectrum is proportional to $g/\Omega_{pqr}$
\citep{Lichtenberg:ChaoticDynamics:82}, thus, the higher the value
of $g$, the larger the number of frequencies that will be present
(remember that Fig.~\ref{fig:Frequencies} displays the frequencies
whose amplitude is above a threshold), and smaller frequencies will
be favored. The intersection of two branches correspond to a resonance,
that is, the condition $\Omega_{pqr}=0$
is fulfilled for some $p,q,r$. Close to a resonance the KAM perturbation
theory breaks down, and non-integrable behaviors appear that may lead
to chaos. In Fig.~\ref{fig:Frequencies} the arrows indicate the
resonances $\Omega_{1q0}=0$ ($q=-2)$ for $g\approx0.093$, and $q=-3,-4,-5$
for, respectively, $g\approx0.14$, $g=0.17,$ $g=0.196$ . For higher
$g$ values, resonances with higher and higher $q$ values become
significative, as seen on the left of the chaotic region: Each frequency
crossing produces a multi-frequency structure (indicated by the arrows). 
Finally, for $g\gtrsim0.2$
one observes at least four successive windows of dense spectrum, corresponding
to a chaotic behavior.

Fig.~\ref{fig:Frequencies} represents a ``local route'' (in
the sense that its detailed geometry depends on the initial conditions)
to quasi-classical chaos. The above argument suggests however that
this route is characteristic of KAM systems, and globally independent of initial
conditions. Its universal character is confirmed by comparing it to
Fig.~\ref{fig:QKR-CKR}a, where an analogous structure is observed
in a completely different KAM system, the classical kicked rotor.
We have thus put into evidence a route to KAM chaos in a nonlinear
quantum system, which is potentially observable with state-of-art
experiments.

\begin{figure}
\begin{centering}
\includegraphics[width=8cm]{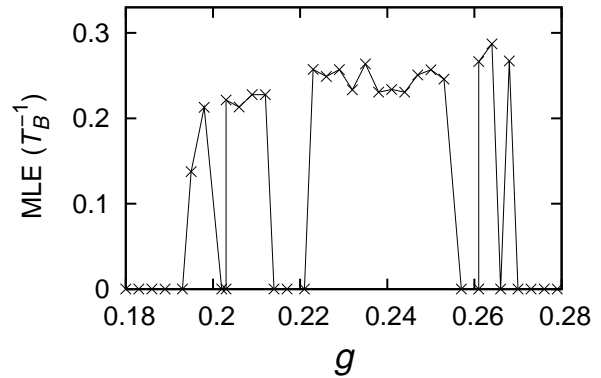}
\par\end{centering}

\caption{\label{fig:Lyapunov}Maximum Lyapunov exponent (in units of $T_{B}^{-1}=\omega_{B}/2\pi$)
as a function of $g$. One observes a good correspondence between
the existence of a non-zero Lyapunov exponent and the presence of
dense-spectrum regions in Fig.~\ref{fig:Frequencies}. The observed
dense-spectrum is thus a reliable signature of the chaotic behavior.
Same parameters and initial conditions as in Fig.~\ref{fig:Frequencies}.}

\end{figure}

In order to confirm our conclusions, we have also calculated the maximum
Lyapunov exponent (MLE) associated to the dynamics, which is a direct
signature of the sensitivity to the initial conditions, and thus of
chaos. In order to calculate MLEs in our quantum system, we adapted
the classical Jacobian method \citep{Ruelle:Chaos:IJMP85}. We represent
the system evolution by a trajectory in a six-dimensional {}``generalized
quantum phase-space'', formed by the real and imaginary parts of
each WS state coefficient, i.e.~$\sqrt{I_{n}}\cos\theta_{n}$ and
$\sqrt{I_{n}}\sin\theta_{n}$, for $n=-1,0,1$ (this is numerically
more stable than using $I_{n}$ and $\theta_{n}$), and calculate
the divergence of neighbor WS states, from which we can extract the
MLE. The result is presented in Fig.~\ref{fig:Lyapunov}. One sees
a clear transition to chaos for $g\approx0.194$, which is in good
agreement with the value that can be deduced from Fig.~\ref{fig:Frequencies}.
We can also identify the four successive chaotic windows.

\begin{figure}
\begin{centering}
\includegraphics[width=8cm]{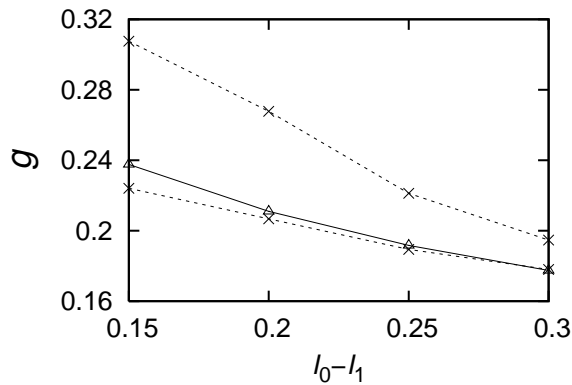}
\par\end{centering}

\caption{\label{fig:gcr}Determination of the critical value of $g$. The solid
lines with triangles are the estimated critical value of $g$, given
by the condition that the trajectory energy equals the separatrix
energy (cf. text). The dotted lines with crosses are the limits of
the chaotic region, where the Lyapunov exponent is non-zero. The parameters
are the same as in Fig~\ref{fig:Frequencies}, with $\theta_{0}-\theta_{1}=\frac{\pi}{2}$.}

\end{figure}

For given initial conditions, we can obtain an estimate for the critical
value of $g$ at which chaos appears. KAM theory shows that chaos
generally appears along a separatrix. In our system there are two
main kinds of trajectories \citep{AP:ChaosBEC:PRL03}: ``Passing''
trajectories correspond to Bloch oscillations slightly perturbed by
the nonlinearity (for the low values of $g$ we are considering);
they appear when the three populations are comparable. Bound, periodic
orbits correspond to a motion essentially confined at a potential
well, and appear when one population dominate the others. Between
these two kinds of trajectories, there is a separatrix. For a given
trajectory (i.e.~for fixed initial conditions),
chaos appears when the changing in the value of $g$ brings it close to the separatrix.
The condition for that can be simply estimated by confining the BEC
to only two wells ($I_{-1}=0$), in which case the system is integrable.
This ``reduced'' system does not exhibit chaos, but it has a phase-space
structure analogous to that of the 3-wells, displaying bound and passing
trajectories, and, between them, a separatrix. In the reduced system
we can calculate both the energy $E_{0}$ corresponding to the trajectory
(with $I_{-1}=0$) and the energy $E_{s}$ of the unstable point
to which the separatrix is connected. We can then numerically determine
the value of $g$ for which $E_{0}=E_{s}$, which gives the critical
value. Fig.~\ref{fig:gcr} shows the result for various initial conditions
(solid line). The dotted lines indicate the limits of the chaotic
region as inferred from the Lyapunov exponent calculation in the 3-well
system. The agreement is very good, even when the chaotic zone is
narrow.

In conclusion, we have characterized the phenomenon of quasi-classical
chaos using an experimentally accessible signal, the mean position
of the boson gas. We have studied a {}``local route'' to quasi-classical
chaos and shown that the information obtained from this approach agrees
well with that provided by a hallmark signature of chaos, the sensitivity
to initial conditions, quantified by a positive Lyapunov exponent.
The understanding of the structure of such a route allowed us to determine
the critical value of the nonlinearity parameter, which is in good
agreement with the one deduced from the calculation of the Lyapunov
exponent. We think that the present work might stimulate an experimental
observation of quasi-classical chaos, which would, in turn, stimulate
new investigations on the nature of quantum chaos.

\begin{acknowledgments}
The authors are happy to thank M. Lefranc for fruitful discussions
and for his help with the calculation of Lyapunov exponents.
\end{acknowledgments}

\end{document}